\documentclass[sigconf,authorversion,nonacm,9pt]{acmart}
\renewcommand\footnotetextcopyrightpermission[1]{} 
\usepackage{fancyhdr}
\usepackage[normalem]{ulem}
\usepackage{url}
\usepackage{tikz}

\usepackage{amsthm}
\usepackage{amsmath,amsfonts}
\usepackage{algorithmic}
\usepackage{graphicx}
\usepackage{textcomp}
\usepackage[table,xcdraw]{}

\usepackage{booktabs} 
\usepackage{stfloats}
\usepackage[normalem]{ulem}
\usepackage{subfigure}
\usepackage{multirow}
\usepackage{paralist}
\usepackage{balance}
\usepackage{bm}
\usepackage[ruled,linesnumbered]{algorithm2e}
\usepackage{diagbox}
\usepackage{threeparttable}
\usepackage{makecell}
\usepackage{colortbl}
\usepackage{bm}
\newcommand{\approach}{MEIC}
\newcommand\name{MEIC}

\usepackage{lipsum}
\usepackage[compact]{titlesec}
\titlespacing*{\section}{0pt}{1.5ex plus .5ex minus .2ex}{1ex plus .2ex}
\titlespacing*{\subsection}{0pt}{1ex plus .2ex minus .2ex}{.5ex plus .1ex}

\DeclareMathDelimiter{(}{\mathopen} {operators}{"28}{largesymbols}{"00}
\DeclareMathDelimiter{)}{\mathclose}{operators}{"29}{largesymbols}{"01}

\definecolor{grey}{RGB}{0.5,0.5,0.5}
\usepackage{tcolorbox}

\newtcolorbox{infobox}[2][]
{
  colframe = grey!25,
  colback  = grey!10,
  coltitle = grey!20!black,  
  title    = #2,
  #1,
}

\newcommand{\parlabel}[1]{{\noindent\bf #1}}
\newcommand{\eg}{e.g.\xspace}
\newcommand{\ie}{i.e.\xspace}
\newcommand{\etc}{etc.\xspace}

\newrobustcmd*\circled[1]{\tikz[baseline=(char.base)]{
            \node[shape=circle,draw,inner sep=1pt,fill,text=white,minimum size=1em] (char) {\textsf{\small #1}};}}

\begin{abstract}
The deployment of Large Language Models (LLMs) for code debugging (e.g., C and Python) is widespread, benefiting from their ability to understand and interpret intricate concepts. 
However, in the semiconductor industry, utilising LLMs to debug Register Transfer Level (RTL) code is still insufficient, largely due to the underrepresentation of RTL-specific data in training sets. 
This work introduces a novel framework, \textbf{Make Each Iteration Count (\name)}, which contrasts with traditional one-shot LLM-based debugging methods that heavily rely on prompt engineering, model tuning, and model training.
\name\ utilises LLMs in an iterative process to overcome the limitation of LLMs in RTL code debugging, which is suitable for identifying and correcting both syntax and function errors, while effectively managing the uncertainties inherent in LLM operations.
To evaluate our framework, we provide an open-source dataset comprising 178 common RTL programming errors. 
The experimental results demonstrate that the proposed debugging framework achieves fix rate of 93\% for syntax errors and 78\% for function errors, with up to 48x speedup in debugging processes when compared with experienced engineers.
The Repo. of dataset and code:
\textbf{\url{https://anonymous.4open.science/r/Verilog-Auto-Debug-6E7F/}}.
\end{abstract}

\begin{document}

\title{ \approach: Re-thinking RTL Debug Automation using LLMs }
\author{Ke Xu$^1$, Jialin Sun$^1$, Yuchen Hu$^1$, Xinwei Fang$^2$, Weiwei Shan$^1$, Xi Wang$^1$ and Zhe Jiang$^1$}
\affiliation{%
  \institution{$^1$National Center of Technology Innovation for EDA, School of Integrated Circuits, Southeast University, China\\$^2$Department of Computer Science, University of York, UK}
  \country{}
}

\maketitle

\thispagestyle{plain}
\pagestyle{plain}

\section{Introduction}
\label{sc:Intro}
In hardware development, the verification and debugging processes are notably laborious and time-consuming, requiring twice the duration of the design phase itself~\cite{lahti2018we}. This significant investment in time and resources shows the need for more efficient methods.

Large Language Models (LLMs) have the potential to revolutionise this process, which have demonstrated a remarkable capability to interpret hardware specifications using natural language and generate corresponding Register Transfer Level (RTL) code, such as Verilog and VHDL. Existing efforts ~\cite{liu2023verilogeval,thakur2023verigen,blocklove2023chip, delorenzo2024make} have showcased the potential of LLMs in automating hardware design, but have also revealed significant limitations. The primary issues are related to the unstable performance of LLMs and the intrinsic complexities of RTL code itself~\cite{zhang2023siren}, which often result in error-prone outputs. 
In response to these limitations, a growing body of research, including RTLFixer~\cite{tsai2023rtlfixer}, SBF~\cite{ahmad2023fixing}, LLM4SecHW~\cite{fu2023llm4sechw}, HDLdebugger~\cite{yao2024hdldebugger}, and AssertLLM~\cite{fang2024assertllm}, has been undertaken to enhance LLM-based RTL debugging. These studies employed techniques such as prompt engineering~\cite{white2023prompt,sahoo2024systematic}, model tuning~\cite{liu2023verilogeval,chang2023chipgpt}, and model training~\cite{liu2023chipnemo,goh2024english} to address these challenges. However, despite these efforts, the performance of these approaches is still far from practical, as evidenced by persistently low \emph{pass@k} rates\footnote{The pass@k metric measures the probability that at least one of the top k outputs generated by a model correctly solves a given problem, used to evaluate the effectiveness of solutions in tasks like code generation and debugging.}.

\begin{figure}[t]
    \centering
    \hspace{5pt}
    \includegraphics[width=1\columnwidth]{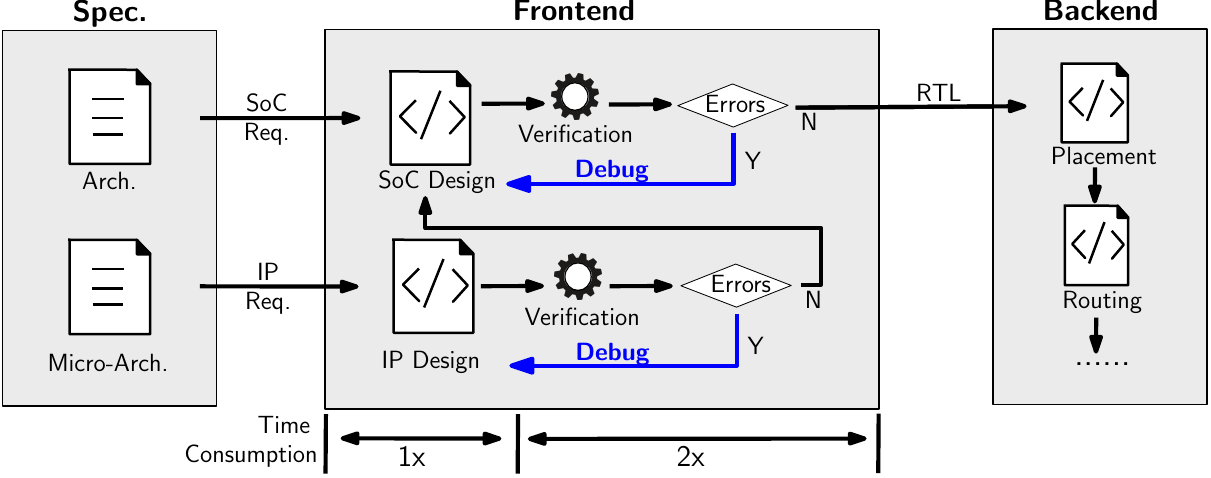}
    \caption{Hardware development flow in the human world \emph{(Spec: Specification; Arch: Architecture; Req: requirement)}: the flow involves the specification definition, frontend development, and backend implementation. 
    After the design requirement is defined, the RTL is coded at both IP and SoC levels. 
    To ensure the design's correctness, multiple iterations of the verification and debugging must proceed, usually consuming twice the duration compared to the design phase.}
    \label{fig:HumanWorld}
\end{figure}

In contrast to previous works, our approach is inspired by established human debugging practices, recognising that \emph{``there is never one-shot debugging''}. 
As depicted in Figure~\ref{fig:HumanWorld}, the debugging process in human-led environments is not only collaborative but also iterative~\cite{wei2023decisive,wei2022designing,laeufer2024rtl,jiang2021bridging,lourencco2023bugdoc}. 
Each stage of the debugging process, from the initial RTL design phase to final verification, involves multiple iterations where different individuals with diverse capabilities engage in verifying and debugging the code. 
This method continues until the code achieves an error-free state or meets stringent coverage criteria. 
Acknowledging that uncertainties in LLM outputs are similar to the variabilities in human performance, this human-led model provides a solid foundation for developing LLM-based RTL debugging methods. 
Particularly, employing an iterative approach addresses the inherent uncertainties associated with LLM models.

\parlabel{Contributions.}
We present \textbf{Make Each Iteration Count (\name)}, a novel framework that utilises multiple LLMs to enable automated and iterative debugging of RTL code. \name\ is designed to address the above limitations commonly associated with applying LLMs in hardware debugging. The main contributions of this paper are:

\vspace{-3pt}
\begin{itemize}
    \item \textbf{An iterative framework:} \name\ integrates RTL toolchain (\eg, compilers and simulators), with two LLM agents, and a code repository. This allows continuous evaluation, testing and debugging of RTL code, mitigating uncertainties caused by the fluctuations in the performance of LLM outputs.
    
    \item \textbf{Dual LLM deployment:} \name\ employs a fine-tuned debug agent that identifies and attempts to correct syntax and function errors, followed by an LLM scoring agent that assesses the quality of RTL candidates derived from the debug agent. This deployment provides quantified and traceable feedback that informs further iterations.
    
    \item \textbf{An open-sourced tooling:} To ease the adoption of  \name, we developed a tooling to enable wide and early utilisation of \name\, which is publicly available on \textbf{\url{https://anonymous.4open.science/r/Verilog-Auto-Debug-6E7F/}}.
    
    \item \textbf{Extensive empirical validation:} We present an open-source error dateset derived from RTLLM~\cite{lu2023rtllm}. This evaluation dataset contains 178 code instances generated from our random error generator which include common syntax and function errors across various modules of combinational and timing logic circuits. 
    
    \item \textbf{Demonstrated performance improvement:} \name, incorporating GPT-4, significantly enhances debugging automation and performance, achieving a syntax error fix rate of 93\% and a function error fix rate of 78\% using our test dataset. Also, it delivers up to 48x speedup in debugging processes when compared with experienced engineers. 
\end{itemize}

\parlabel{Organisation.} The rest of the paper is organised as follows: Section~\ref{sc:Concepts} presents the top-level concepts of \name, Section~\ref{sc:Details} introduces the details of \name\ and the rationales.
Section~\ref{sc:Evaluation} evaluates our framework against five research questions, followed by the related work given in Section~\ref{sc:RelatedWork}. 
Section~\ref{sc:Conclusion} concludes and offers the insights.


\vspace{-2.5pt}
\section{\name: An Overview}
\label{sc:Concepts}
\vspace{-2.5pt}

\begin{figure}[htbp]
    \centering
    \vspace{-10pt}
    \includegraphics[width=1.01\columnwidth]{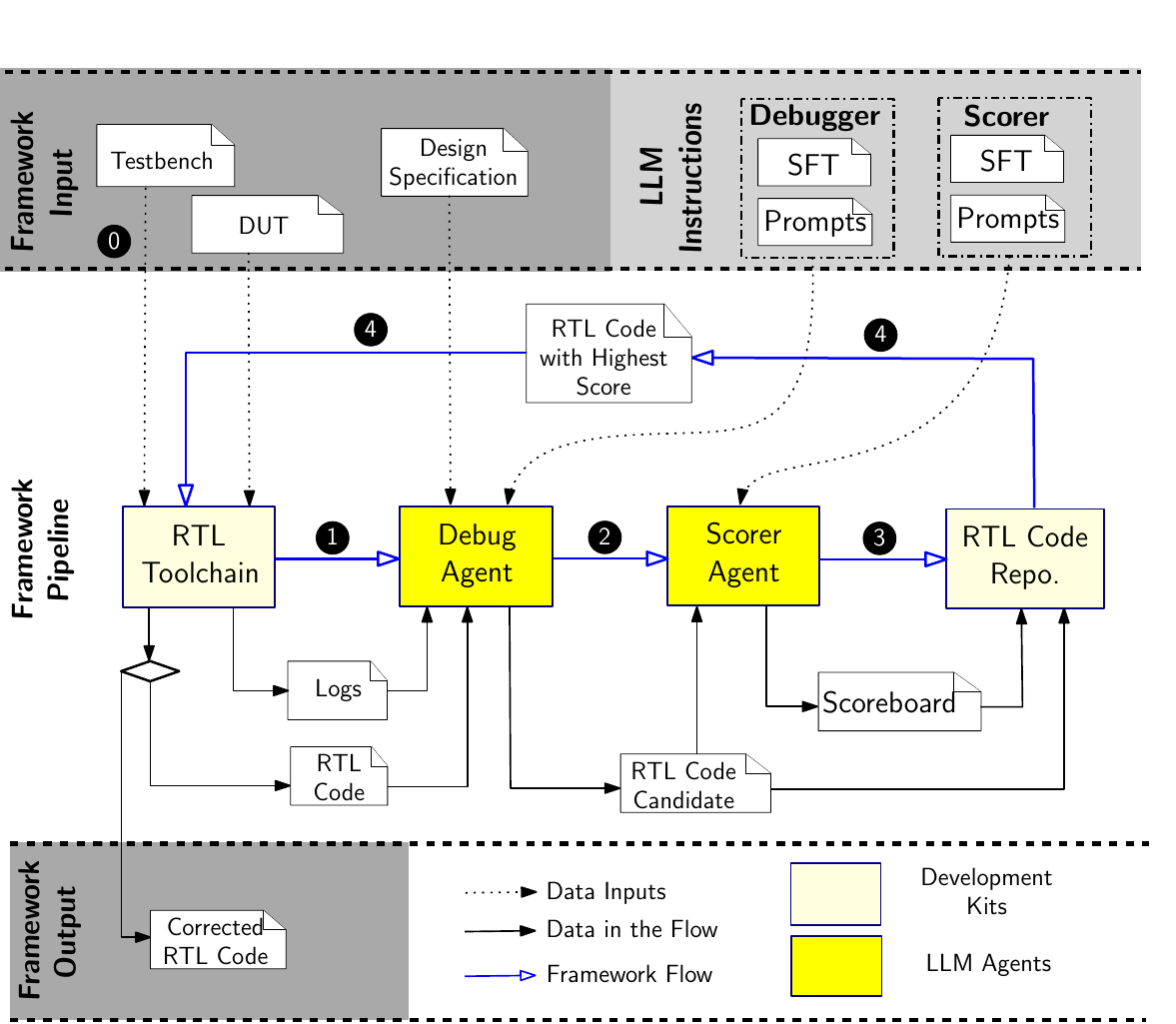}
    \caption{\name\ overview:
    the framework initialises with the DUT, which is compiled and simulated by the RTL toolchain (step \circled{0}).
    The resultant logs and code are forwarded to the debug agent for error resolution (step \circled{1}). 
    The revised RTL code is examined by the scorer agent (step \circled{2}) and stored in the repository (step \circled{3}), from which the highest-scored code is selected for the following debugging iteration (step \circled{4}).
    }
    \hspace{-15pt}
    \label{fig:overview}
\end{figure}

Intended for use in the design and verification stages, \name\ aims to help hardware developers identify and correct both syntax and function errors in RTL code. 
This systematic framework (see Figure~\ref{fig:overview}), wraps a RTL toolchain (e.g., compilers and simulators), two LLM agents (fine-tuned\footnote{We acknowledged that fine-tuning's definitions are various in different LLMs. Here we use the GPT4 as an example: \url{https://platform.openai.com/docs/guides/fine-tuning}.}  for code debugging and assessment), and a code repository. To ensure the framework's applicability across different scenarios, we standardised  its inputs as:

\begin{itemize}
\vspace{-1.5pt}
    \item \textbf{Design specification}: outlining the intended and expected behaviour of the hardware component;
    
    \item \textbf{RTL code}: containing the untested RTL code of the initial hardware design, \ie, Design Under Test (DUT).
        
    \item \textbf{Testbench}: acting as the reference for verifying the functional correctness of the RTL code.
\end{itemize}
\vspace{-1.5pt}

\approach\ assumes that the LLMs employed can perform better through proper fine-tuning and prompt engineering. Our approach for this processing (e.g., fine-tuning and prompt engineering) of LLMs is discussed in Section \ref{sc:Details}.
Under these assumptions, \approach\ attempts to correct the RTL code as necessary across a number of iterations from four pipeline stages (Step \circled{1}-- \circled{4}) in Figure~\ref{fig:overview}, highlighting our principle that debugging is an iterative, not an one-shot process. 
The iterative \approach\ pipeline involves the following steps:

\begin{itemize}
    \item \textbf{Step \circled{0}:} \approach\ begins by taking the user's inputs, which are processed in the compiler and simulator to detect the syntax and function errors, respectively. 
    If no errors are found, the process terminates, outputting the error-free code. 
    If errors are detected, the code, its associated logs, and the design specifications are sent to the debug agent;
    
    \item \textbf{Step \circled{1}:} the debug agent is expected to correct the erroneous RTL code (both syntax and function errors) by producing a code candidate based on its inputs;

    \item \textbf{Step \circled{2}:} the RTL code candidate is analysed and evaluated by the scorer agent, which assesses the quality of the generated code candidate and assigns a numerical score;

    \item \textbf{Step \circled{3}:} the RTL code candidate and its score are stored in the repository to enable a rollback mechanism, preventing the current RTL code from being overwritten by potentially incorrect LLM outputs, \eg, skipping lines of the code;

    \item \textbf{Step \circled{4}:} the repository selects the RTL code with the highest score for the new interaction round, continuing until the framework meets the termination condition.
\end{itemize}

The framework terminates and outputs the latest design code analysed in the toolchain if \textbf{i)} no error is found or \textbf{ii)} it reaches the threshold of the maximum number of iterations. 

\parlabel{Modularisation and flexibility.}
It is worth noting that standard interfaces between different stages of the pipeline are used. This ensures easy upgrades and extensions. For instance, replacing the debug agent with a domain-specific model or altering the RTL toolchain can be achieved by simply modifying the API or maintaining consistent log formats. This flexibility allows \name\ to adapt to various debugging scenarios and technological advancements.


\parlabel{Contexts.}
As a framework, \name\ is agnostic to the RTL and LLMs. 
For illustrative purposes, we used Verilog, the most widely adopted RTL language in the industry, along with its associated toolchain, \emph{ModelSim}~\cite{modelsim}, as an example throughout the paper.
For LLMs, we used different versions of GPT models from OpenAI (see Section~\ref{sc:Evaluation}).

\vspace{-5pt}
\section{\name: The Framework Pipeline}
\label{sc:Details}
\vspace{-3pt}
We first discuss our preparations before the operation of the pipeline, including the error classifications (Section 3.1) and the tuning method (Section 3.2), both of which aim to enhance the LLM's performance. For the operation of the pipeline, we introduce the simulation process that uses the RTL toolchain combined with feedback engineering (Section 3.3), followed by the discussion on the microsystems integrated with the LLM agents for debugging (Section 3.4) and for the best-version code selection (Section 3.5). Finally, we briefly introduce the proposed open-source Verilog error dataset (Section 3.6) that is used in the evaluation.





\begin{table*}
    \sffamily
    \small
    \centering
    \vspace{-25pt}
     \caption{Common Verilog error categories and examples.} 
     \vspace{-2pt}
     \label{Table:VerilogErrorsV2}
    \resizebox{0.99\linewidth}{!}{%
    \begin{tabular}{p{0.6cm} p{4cm}<{\centering} p{8cm}<{\centering}  >{\columncolor{green!30}}p{4.9cm}<{\centering}  >{\columncolor{red!30}}p{4.9cm}<{\centering}} 
    \toprule
    \rowcolor{white} \textbf{Types} & \textbf{Error} & \textbf{Description}& \textbf{Expected Form} & \textbf{Unexpected Form} \\ 
    \midrule
    \multirow{15}{*}{\rotatebox[origin=c]{90}{\parbox[c]{20mm}{\centering \textbf{Syntax Errors}}}}& \makecell[l]{Premature Termination} & \makecell[l]{Missing or redundant punctuation (e.g., semi-colons or commas)  \\causing premature end of the execution.}&\makecell[l]{module A(input a, \textbf{output b);}}&\makecell[l]{module A(input a, \textbf{output b)}}\\ \cmidrule{2-5}
    & \makecell[l]{Undefined Variable} &  \makecell[l]{Using variables that have not been previously declared.}&\makecell[l]{assign \textbf{result} = temp;}&\makecell[l]{assign \textbf{resutl} = temp;} \\ \cmidrule{2-5}
    & \makecell[l]{Operator Misuse} & \makecell[l]{Operator misuse (e.g., incorrectly using the assignment operator `=' \\instead of the comparison operator `==' for evaluating conditions) \\resulting in unacceptable expression format.}&\makecell[l]{if (\textbf{a} \bm{$==$} \textbf{2'b10}) \\begin b <= 1'b1; end}&\makecell[l]{if (\textbf{a} \bm{$=$} \textbf{2'b10}) \\begin b <= 1'b1; end} \\ \cmidrule{2-5} 
    & \makecell[l]{Redundant Variable Declaration} & \makecell[l]{Declaring the same variable multiple times e.g. in port definitions.}&\makecell[l]{module A(input a, output b);\\ reg \textbf{a\_temp};}&\makecell[l]{module A(input a, output b);\\ reg \textbf{a};} \\ \cmidrule{2-5}
    & \makecell[l]{Incorrect Encoding}& \makecell[l]{Presence of characters not aligning to ANSI encoding standard.}&\makecell[l]{module A(input \textbf{a}, output b);}&\makecell[l]{module A(input \textbf{â}, output b);}\\ \cmidrule{2-5}
    & \makecell[l]{Incorrect Data Type Assignment} & \makecell[l]{Failing to comply with assignment rule for reg- and wire-type data.}&\makecell[l]{reg a;\\always @(*) begin \textbf{a} \bm{$=$} \textbf{b}; end}&\makecell[l]{reg a;\\ \textbf{assign a} \bm{$=$} \textbf{b};} \\ \cmidrule{2-5}
    & \makecell[l]{Port Mode Declaration Error}& \makecell[l]{Failing to declare module port according to the rules.}&\makecell[l]{module A(a, b);\\ input a;\\ \textbf{output b;}}&\makecell[l]{module A(a, b);\\input a;\\  \textbf{//Declaration for b is missing.}} \\ \cmidrule{2-5}
    & \makecell[l]{Data Index Out-of-Bounds Error} & \makecell[l]{Exceeding allowable data bounds during array or vector operations.}&\makecell[l]{reg [32:1]a;\\ assign b = \textbf{a[16:1]};}&\makecell[l]{reg [32:1]a;\\ assign b = \textbf{a[15:0]};} \\ \cmidrule{2-5}
    & \makecell[l]{Improper Use of Keywords} & \makecell[l]{Using reserved keywords incorrectly or as identifiers.}&\makecell[l]{reg \textbf{alway};}&\makecell[l]{reg \textbf{always};} \\ \midrule
        \multirow{15}{*}{\rotatebox[origin=c]{90}{\parbox[c]{40mm}{\centering \textbf{Function Errors}}}}& \makecell[l]{Insufficient Bit Width} &  \makecell[l]{Defining registers with inadequate bitwidth.}&\makecell[l]{wire \textbf{[3:0]} a;\\assign a = 4'b1000;}&\makecell[l]{wire \textbf{[3:1]} a;\\assign a = 4'b1000;} \\ \cmidrule{2-5}
    &\makecell[l]{Incomplete Port Connection} & \makecell[l]{Failing to connect all ports during module instantiation in Verilog.}&\makecell[l]{mod md(.a(a), \textbf{.b(b)});}&\makecell[l]{mod md(.a(a), \textbf{.b()});} \\ \cmidrule{2-5}
    & \makecell[l]{Flawed Sensitivity List}& \makecell[l]{Omitting or mis-specifying signal data in Verilog's sensitivity list.}&\makecell[l]{always @(posedge clk or \textbf{negedge} rst\_n) \\  begin a <= b + c; end}&\makecell[l]{always @(posedge clk or \textbf{posedge} rst\_n) \\begin a <= b + c; end} \\ \cmidrule{2-5}
    & \makecell[l]{Misuse of Assignments} & \makecell[l]{Misusing blocking (=) and non-blocking (<=) in sequential design.}&\makecell[l]{always @(posedge clk or negedge rst\_n) \\  begin \textbf{a} \bm{$<=$} \textbf{b} \bm{$+$} \textbf{c}; end}&\makecell[l]{always @(posedge clk or negedge rst\_n) \\  begin \textbf{a} \bm{$=$} \textbf{b} \bm{$+$} \textbf{c}; end} \\ \cmidrule{2-5}
    & \makecell[l]{Logical Errors in Expressions} &  \makecell[l]{Complex and incorrect module logic during code formulation.}&\makecell[l]{assign a = \textbf{b} \bm{$+$} \textbf{c};}& \makecell[l]{assign a = \textbf{b} \textbf{\&} \textbf{c};}\\ \cmidrule{2-5}
    & \makecell[l]{Concurrent Variable Use} & \makecell[l]{Assigning the same variable in multiple processes.} &\makecell[l]{always @(*) begin a=\textbf{1'b1}; end}&\makecell[l]{always @(*) begin a=\textbf{1'b1}; end\\always @(*) begin a=\textbf{1'b0}; end}\\ \cmidrule{2-5}
    & \makecell[l]{Mismatched Assignment Values} &  \makecell[l]{Omitting base indication in values that leads to unexpected assign-\\ments.}&\makecell[l]{if (a == \textbf{2'b10}) \\begin b <= 1'b1; end}&\makecell[l]{if (a == \textbf{10}) \\begin b <= 1'b1; end} \\ \cmidrule{2-5}
    & \makecell[l]{Incorrect Module Instantiation} & \makecell[l]{Instantiating a non-existent module, but only fails in functionality.}&\makecell[l]{\textbf{mod} md(.a(a), .b(b));}&\makecell[l]{\textbf{mdo} md(.a(a), .b(b));} \\ \cmidrule{2-5}
    & \makecell[l]{Infinite Loop Constructs} & \makecell[l]{Loops using \textit{forever}, \textit{while}, or \textit{for} without a clear termination condi-\\tion will not end.}&\makecell[l]{next\_stage <= \textbf{next\_stage\_temp};}&\makecell[l]{next\_stage <= \textbf{current\_stage};} \\ \bottomrule
    \end{tabular}
    }
    \vspace{1.5pt}
    \label{error_c}
\end{table*}

\vspace{-2.5pt}
\subsection{Error Classifications}
\label{sbsc:error}
\vspace{-2.5pt}

Understanding error classifications in Verilog is essential for effective debugging. For human engineers, knowing whether an issue is a syntax or function error allows for the selection of appropriate tools and techniques (compilers and linters for syntax errors, and simulators, waveform analysers, and timing analysis tools for function errors). Similarly, for LLMs, this classification would aid in executing more accurate debugging processes as it would allow for a more structured reasoning~\cite{wei2022chain,nam2024using, cohn2024chain,wu2022ai}. 


\parlabel{Syntax errors.}
Syntax errors are errors that occur when the code violates the formal structure of the Verilog language. 
Such errors are typically identified by compilers during the parsing stage, preventing further simulation or synthesis. The compiler, \eg, ModelSim and DC, produces logs detailing the location and nature of these errors, thus helping with quick error detection and correction.

\parlabel{Function errors.}
Function errors encompass all other errors that affect the operation code and include semantic errors, logical errors, timing errors, \etc 
Unlike syntax errors, function errors are concerned with the behaviour and outcome of the code rather than its grammatical correctness. 
These errors can be more challenging to detect and often require extensive testing, simulation, formal verification, and detailed examination of timing and synthesis reports.
For practical identification of such errors, assertions and testbench are commonly employed to identify unexpected outcomes for subsequent correction (see Section~\ref{sbsc:simu}).

Drawing on the practical errors identified in our past hardware designs and insights from previous studies \cite{chonnad2007verilog, sutherland2010verilog}, we categorised the Verilog errors and gave corresponding examples in Table \ref{Table:VerilogErrorsV2}.

\begin{figure}[t]
    \centering
    \includegraphics[width=1.00\columnwidth]{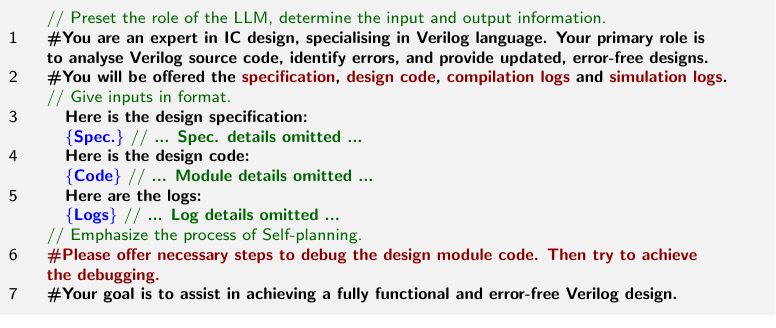}
    \vspace{-0.35cm}
    \caption{Part of the input patterns for the self-planning. 
    In addition to debugging based on provided files (lines 1-4), the agent is also required to plan the debugging process (line 5).}
    \label{fig:spprm}
\end{figure}

\vspace{-4pt}
\subsection{Tuning Method}
\label{sbsc:ft}
\vspace{-3pt}
\parlabel{Domain-specific knowledge.} 
It is well-known that LLMs can achieve better performance in a specific domain by utilising system-level instructions and extra domain knowledge~\cite{fu2023llm4sechw,liu2023chipnemo,pal2024domain}.
Based on this property, we fine-tuned the LLM by supplying relevant information as system-level instructions and incorporating domain knowledge through the prompts~\cite{gpt-tuning,zamfirescu2023johnny,wan2024software,srikumar2023fast,feng2024don}.

\begin{figure}[t]
    \centering
    \includegraphics[width=1.00\columnwidth]{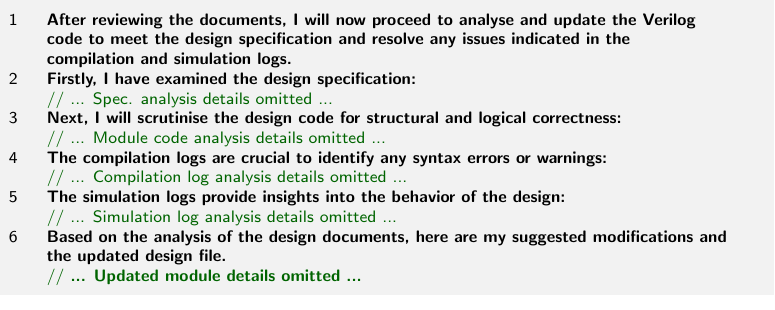}
    \caption{Reply of the LLM. With self-planning technique, the LLM provides a set of steps for RTL analysis and debugging.}
    \label{fig:spprmrp}
\end{figure}

To facilitate the interpretation and correction of Verilog code, we incorporated the Verilog-2001 standard~\cite{sutherland2000ieee} into the knowledge, along with a number of RTL code examples (\eg, HDLBits~\cite{hdlbits}, Verilog-G~\cite{thakur2023benchmarking}, RTL Coding Guidelines~\cite{ramachandran2007rtl}, \etc) and the error knowledge (in Table~\ref{Table:VerilogErrorsV2}). The previously gathered debugging-related data, subjected to cleansing and extraction, given through prompts.

\parlabel{Prompting techniques.}
To further improve the performance of the LLMs in code generation and debugging, we utilised advanced prompt engineering techniques, \eg, \emph{self-planning} and \emph{role prompting} as described in~\cite{lu2023rtllm,jiang2023selfplanning}. The use of self-planning helps the LLMs to break down a complex task into several planned steps and the use of \emph{role prompting} enables more relevant and contextually appropriate responses, allowing structured and relevant LLM responses.

We captured snapshots of our prompts when fine-tuning the LLM, as shown in Figure~\ref{fig:spprm}. The response of the LLM, as presented in Figure~\ref{fig:spprmrp}, provides a clear outline of the five structured steps (i.e., from line 2-6). It suggests that the specification is analysed first, followed by the analysis of the design code, the compliance logs and the simulation logs. Finally, it suggests the RTL modifications.

\subsection{Toolchain, Compilation, and Simulation}
\label{sbsc:simu}
We employed a toolchain to verify RTL code compliance with design requirements by evaluating compilation and simulation results.


As mentioned in Section~\ref{sbsc:error}, syntax errors can be directly detected during the compilation and detailed in the compilation logs, while function errors often go undetected during compilation and result in unexpected outputs during simulation.
Hence, building test cases is crucial for automatically identifying and correcting function errors, ensuring that the output meets expectations.

For Verilog, function errors are typically identified using two common methods: testbenches and assertions.
Because the testbenches and assertions provide different granularity of error information, we integrated both methods in \name.

\begin{figure}[htbp]
    \centering
    \vspace{-5pt}
    \includegraphics[width=1.00\columnwidth]{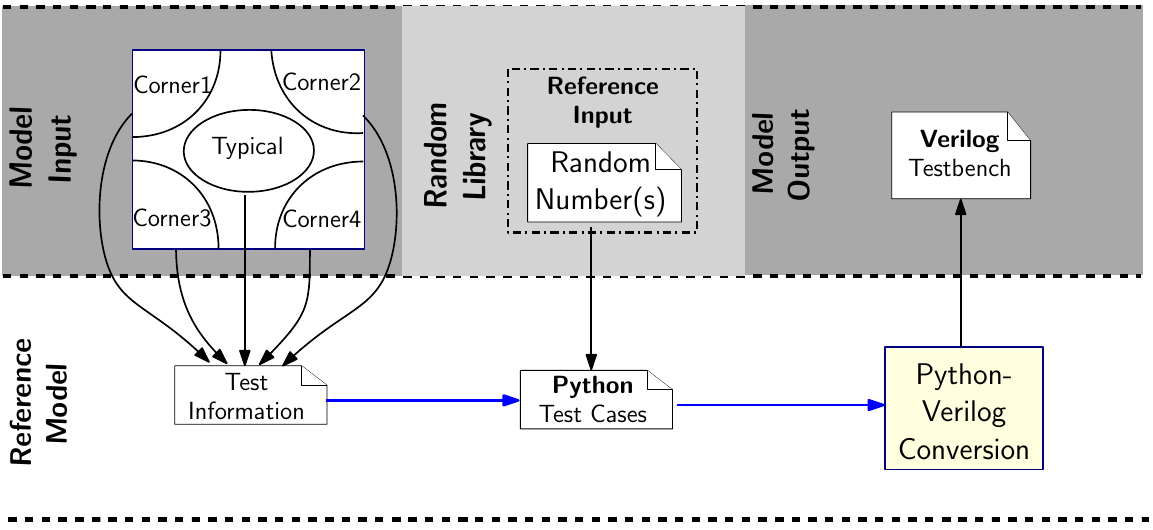}
    \caption{Testbench from reference model to Verilog.}
    \vspace{-5pt}
    \label{fig:gref}
\end{figure}
\begin{figure}[htbp]
	\centering
        \vspace{-5pt}
	\subfigure[Testbench.]{
		\begin{minipage}{\columnwidth} 
                \centering
                \includegraphics[width=0.95\textwidth]{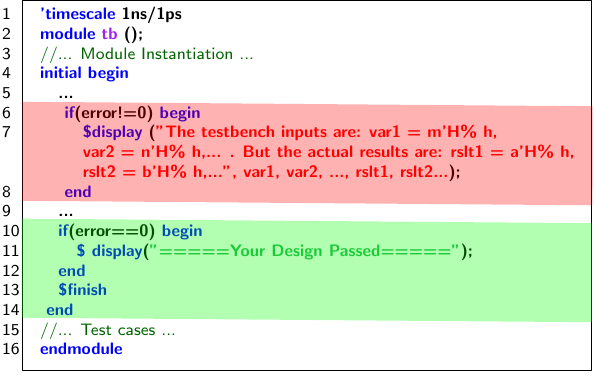} \\
                \label{TB}
		\end{minipage}
	}
	\subfigure[Assertion.]{
		\begin{minipage}{\columnwidth}
                \centering
			\includegraphics[width=0.95\textwidth]{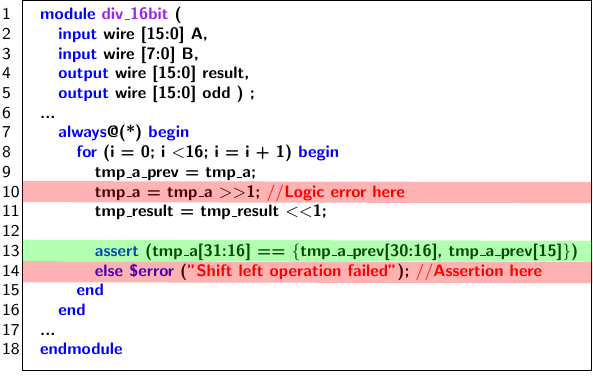} \\	
   \label{SVA}
		\end{minipage}
	}
    \vspace{-5pt}

   \caption{Methods for function errors' detection, which provides different granularity of error information.} 
     \label{fig:dfe}
\end{figure}

\parlabel{Testbench-based detection.}
The testbench serves as a reference model, continuously providing stimulus to the DUT and verifying its outputs against the expected results.
To mitigate common-cause errors between the DUT and the testbench, we developed the test cases in the testbench using Python.
Specifically, we employed \texttt{Random} library to generate input data and wrote the corresponding functionality given in the specification.
With that, we developed an automated script shown in Fig~\ref{fig:gref} that translates the reference model into Verilog syntax, maintaining alignment with the standards.
The \texttt{\$display()} function is used to report the results (Figure~\ref{TB}).

\parlabel{Assertion-based detection.} 
To provide more traceable error feedback, assertions are employed in the conventional debugging approach. 
The original Verilog code is transformed into SystemVerilog code, and assertions are incorporated into the code for error-prone areas. 
If the function at the assertion is erroneous at simulation, the simulation will be terminated directly, and logs will be displayed  (Figure~\ref{SVA}).
Different from the testbench-based detection, which only records comparisons between inputs and outputs, the assertion-based approach offers more precise about the errors.

Following the acquisition of the RTL code and different detection techniques, ModelSim is used for local simulation. 
To ensure the smooth operation of the entire framework, the simulation process is automated to reduce the needs for manual intervention. This is achieved through \texttt{CMD} commands that automate the simulation while generating compilation and simulation logs. 

\parlabel{Other detection techniques.}
We acknowledged that alternative detection techniques are being explored within the community. For instance, references~\cite{blocklove2023chip,lu2023rtllm,hassanllm} demonstrate the use of LLMs to generate testbenches. As outlined in Section~\ref{sc:Concepts}, \name\ is designed to be versatile, supporting a broad ranges of error detection approaches, including those based on LLMs. This compatibility only requires that these techniques supply formalised simulation logs, which are then used as input for the framework. 

\subsection{Debug Agent}
\label{sbsc:de}
The LLM agent functions as an expert in RTL debugging, expecting four inputs: specification, Verilog code, compilation logs, and simulation logs. 
The specification includes a design description that outlines the code's functionality as well as its inputs and outputs.
Following the RTL simulation, both compilation and simulation logs are generated (Section~\ref{sbsc:simu}). 
Using these logs, it is expected that the debug agent can locate the error and leverage its expansive knowledge base (Section~\ref{sbsc:error} and~\ref{sbsc:ft}), giving suggestions for modifications and providing an updated version of the code.


\parlabel{Debug iteration(s).}
Based on the knowledge outlined in Section~\ref{sbsc:ft}, the debug agent then returns the modified design file. However, in most cases, neither human engineers nor LLM agents can get the code right on the first attempt. Therefore, we followed the debugging process of human engineers and introduced an iterative process. If the agent does not generate the correct code in the current iteration, the next iteration will be performed using the highest-scored code from the previous iterations. 

\begin{figure}[t]
    \centering
    \vspace{-30pt}
    \includegraphics[width=1.00\columnwidth]{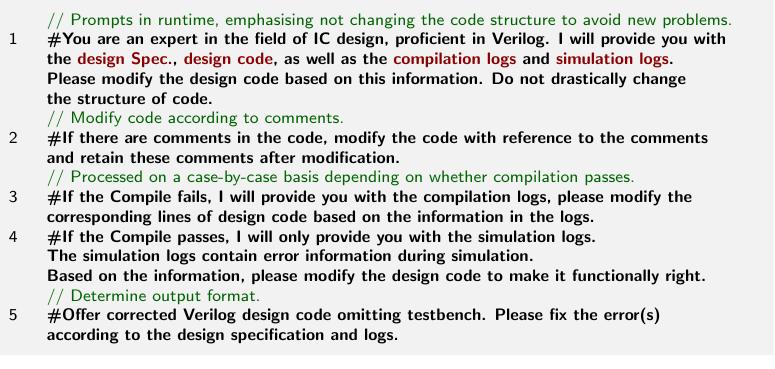}
    \vspace{-14.5pt}
    \caption{System prompts in runtime for the debug agent.}
    \label{fig:spld}
\end{figure}

Furthermore, the type of error encountered dictates the error messages generated. For example, syntax errors only result in the generation of compilation logs, as simulation logs are generated when the syntax is correct. Therefore we need to perform format control based on these two situations. 
The key prompts for the debug agent are shown in Figure \ref{fig:spld}.

\begin{figure}[htbp]
    \centering
    \vspace{-5pt}
    \includegraphics[width=0.95\columnwidth]{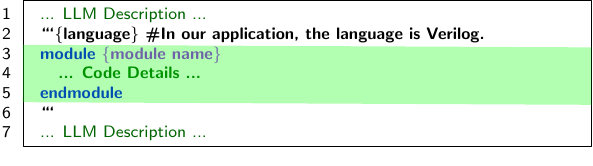}
    \vspace{-5pt}
    \caption{Code output format from the LLM.}
    \vspace{-10pt}
    \label{fig:cofl}
\end{figure}

\parlabel{Formalising agent's outputs.}
 It is often observed that LLM's response often containing irrelevant information to code debugging \eg, the LLM's debugging reasoning process or explanations of the code. Based on our observation, the LLM's code outputs typically follow a specific format, as shown in Figure~\ref{fig:cofl}. To prevent noisy inputs from being used in subsequent iterations, we systematically clean and extract the LLM's responses to ensure that only the essential parts (i.e., the code as shaded in Figure~\ref{fig:cofl}) are carried forward for use in the next iteration.
 


\vspace{-2.5pt}
\subsection{Scorer Agent and Exception Handling}
\vspace{-2.5pt}
\label{sbsc:ehr}

\begin{figure}[htbp]
	\centering
        \vspace{-35pt}
	\subfigure[Code missing.]{
		\begin{minipage}{\columnwidth} 
                \centering
                \includegraphics[width=0.95\textwidth]{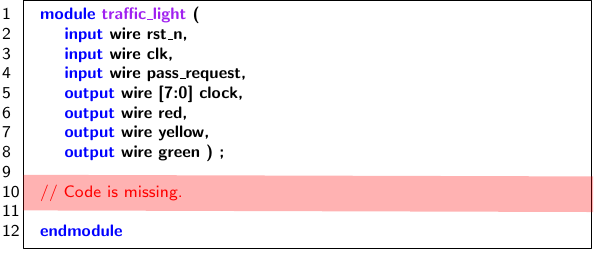} \\
		\end{minipage}
	}
	\subfigure[New error generated.]{
		\begin{minipage}{\columnwidth}
                \centering
			\includegraphics[width=0.95\textwidth]{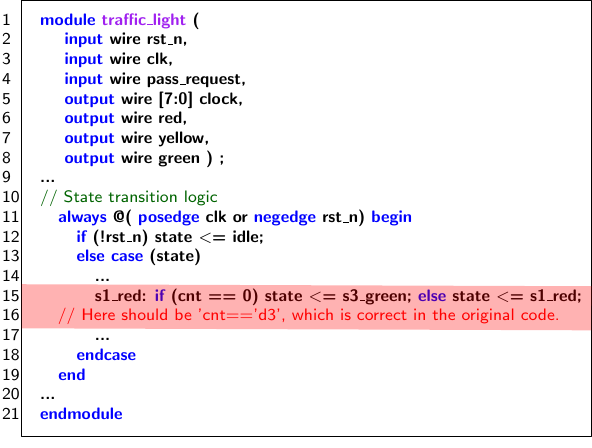} \\			
		\end{minipage}
	}
        \vspace{-5pt}
	\caption{Exceptions should be handled by the scorer agent.} 
	\label{fig:exp}
\end{figure}

While employing the LLMs, unexpected situations may arise, also known as ``hallucination'' ~\cite{ji2023towards,amatriain2024measuring,galitsky2023truth,chen2023hallucination}.
For example, the LLM may return incomplete code.
If subsequent iterations are based on these flawed outputs, the quality of the results may be compromised. Also, the LLM can inadvertently modify both erroneous and correct portions of the code, leading to a situation where most of the iterations are spent addressing new errors introduced by the LLM itself. 
Two common ``exceptions'' are shown in Figure \ref{fig:exp}.
To avoid these exceptions, we introduce \emph{Scorer agent} and \emph{Rollback mechanism} as the exception handling mechanisms in \name.

\parlabel{Scorer agent.}
We introduced a scorer agent to detect unexpected cases.
During the debugging process, modifications should aim to minimise the change to the original code. 
After the debug agent proposes corrections, the modified code is compared to the version from the previous iteration. 
The scorer agent then evaluates the code based on its completeness and overall quality.
If the assessed score falls below a predefined threshold, a rollback mechanism is activated to revert the design code to the most recent comparable and simulatable version. 
To improve the reliability of the scoring, the scorer agent is prompted with a variety of metrics:

\begin{itemize}
    \item 	\textbf{Readability:} The clarity of the code that can be understood.
    \item 	\textbf{Maintainability:} The ease with which the code can be updated or altered.
    \item 	\textbf{Robustness:} The capacity to handle errors and anomalies.
    \item 	\textbf{Standards Compliance:} The alignment of the code with established Verilog coding standards.
\end{itemize}

Although these metrics are qualitative, the scoring process remains effective because all RTL codes are evaluated using the same scorer, ensuring consistency between assessments. 
In addition, the \name\ only interests the relative scores between the iterations.
To further mitigate the uncertainty of the qualitative scoring, a low temperature\footnote{A hyper-parameter in GPT models that controls the randomness of GPT's responses. A lower value is associated with less randomness in their responses} is configured to the scorer agent. 





\parlabel{Rollback mechanism.}
Within the scorer agent, we introduced a mechanism to support possible rollback. This is achieved by saving each version of the Verilog code, along with its corresponding compilation and simulation logs from each iteration, in a designated location. This mechanism not only enables rollback but also enhances traceability as the iteration evolves. 


\begin{figure}[h]
    \centering
    
    \includegraphics[width=1\columnwidth]{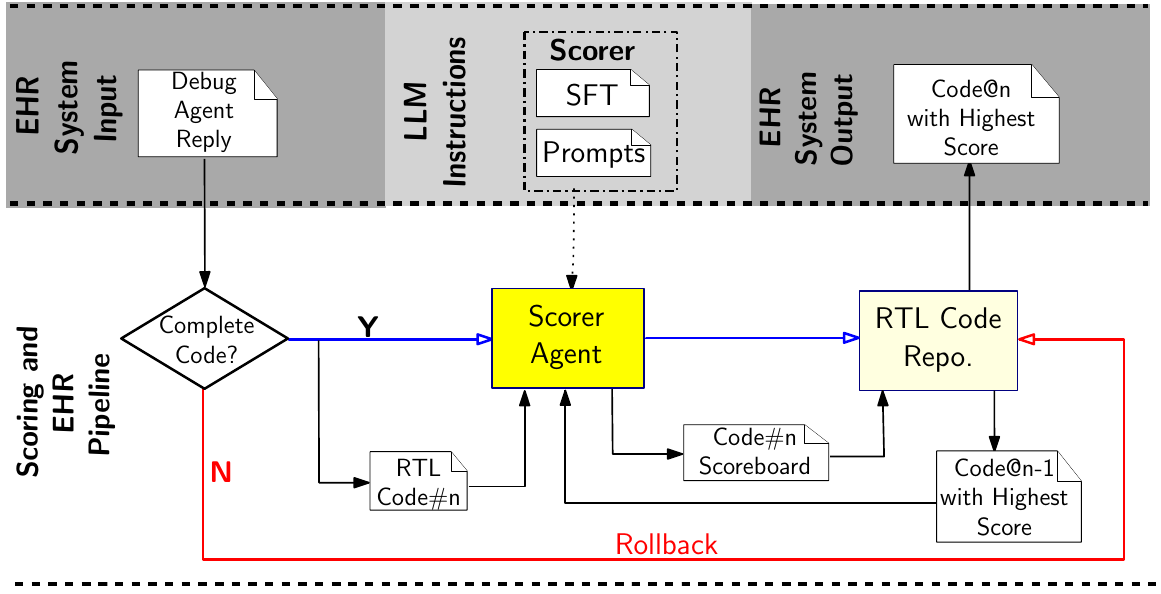}
    \vspace{-5pt}
    \caption{Flow of Exception Handling and Rollback (EHR).}
    \vspace{-10pt}
    \label{fig:ehr}
\end{figure}

The rollback mechanism is primarily triggered based on the completeness of the code produced by the LLM and the score provided by the scorer agent. We assume that the completeness is indicated by the number of lines in the code. If a significant reduction in lines or missing code is observed, we trigger the rollback mechanism regardless of the scorer agent. Otherwise, rollback occurs when the code scored below a predetermined threshold in the scorer agent. Once triggered, the highest-scored code from the previous iteration is used for the next iteration, rather than the most recently produced code. This handling of exception is illustrated in Figure~\ref{fig:ehr}. 







\subsection{Error Dataset}
\label{sbsc:data}

Different modules may exhibit diverse error distributions. To evaluate the debugging capabilities of \approach, we used an open-source dataset~\cite{lu2023rtllm} and intentionally introduced a variety of common errors to construct a specialised dataset tailored for debugging. 
Our goal is to comprehensively represent the spectrum of prevalent Verilog errors by selecting a representative sample of designs.

Because our dataset is developed based on RTLLM~\cite{lu2023rtllm}, some error types might not be generated in certain modules, such as the infinite loop construct in the loop-free modules.
Therefore, this error dataset primarily includes the errors that are prevalent across most modules (summarised in Table~\ref{tab:dataset}), such as premature termination, undefined variable, \etc

\begin{table}[htbp]
    \centering
    \vspace{-5pt}
    \caption{{Verilog error dataset.}}
     \resizebox{0.8\linewidth}{!}{%
    \begin{tabular}{p{11.5em}| p{3.3em}| p{3.3em}|p{3.3em}}
         \toprule
         \diagbox{Module type} {Error type} & Syntax  & Function&Total \\
         \midrule
         Arithmetic & 57 & 38&95 \\
         \midrule
         Logic & 50 & 33&83\\
         \midrule
         Total & 107 & 71&178\\
         \bottomrule
    \end{tabular}}
    \label{tab:dataset}
\end{table}
\vspace{-5pt}



\parlabel{Random error generation.} Given that the majority of the errors introduced are common errors, we also implemented an automatic error generation method, which could be used for more compressive evaluation.
For simpler error types (\eg, misuse of assignments), we devised a set of bug-pattern lists, using regular expressions to identify segments suitable for random error insertion. 
For more complex function errors (\eg, infinite loop), error generation is facilitated through GPT-4 or manual insertion, followed by a data cleansing process. The final dataset size could potentially expand to 200 times the original data size.

\section{Evaluation}
\label{sc:Evaluation}
This section presents our experimental setup, research questions, evaluation metrics and results to answer the questions. 

\parlabel{Setup.} In our experiment, the LLM agents powered by the OpenAI website interface were utilised. Unless otherwise stated, we used the GPT-4 Turbo model as our default LLM agent for both debugging and scoring. We set the temperature of the agent, which controls the randomness of the LLM's output, to 0.7 for the debug agent as deemed optimal in our evaluation of \textbf{RQ1}, and 0.1 for the scorer agent to minimise the randomness of each scoring process as discussed in Section~\ref{sbsc:ehr}. We then developed test cases containing various Verilog design scenarios using ModelSim SE 10.7 simulation environment. We set the threshold of iterations to 10, as based on our experiences, the improvement is hardly observed after that. 



\subsection{Research Questions}
We carried out the experiments to evaluate our framework against five key Research Questions (RQs):

\noindent \textbf{RQ1 (Sensitivity): How does \textit{temperature} in GPT-4 setting impact the performance of \approach\ in terms of FR?} This research question explores the effect of LLM's temperature which controls randomness and lowering the temperature results in less random completions. It seeks to understand how the randomness of LLM's output impact the dubugging performance\footnote{We focused on the debug agent because we argued that the debug agent plays a more important role in debugging workflow than the scorer agent.}. 

\noindent \textbf{RQ2 (Effectiveness): Can \approach\ correct different types of errors across various modules?} This research question investigates the performance of the \approach\ system in terms of its ability to fix errors across a variety of code modules, focusing on both syntax and function errors. It seeks to understand how the system's debugging effectiveness varies depending on the complexity of the code and the type of error encountered.

\noindent \textbf{RQ3 (Impactability): How do various LLM-based configurations and integration impact debugging performance in terms of fix rate?} This research question explores the potential improvements in error correction capabilities through both fine-tuning the models and integrating them with the \approach\ framework. It aims to evaluate whether the proposed \approach\ framework can significantly enhance RTL debugging performance.

\noindent \textbf{RQ4 (Usability): How does \name\ work with different LLM agents?} This research question aims to compare the effectiveness of integrating different LLM models in \name, specifically GPT-3.5 and GPT-4, in debugging code. It seeks to understand how different LLM models influence the performance of \name, quantifying their usability for RTL debugging. 
 

\noindent \textbf{RQ5 (Performability): How does \approach\ compare with human experts in debugging performance?} This research question evaluates how the \approach\ debugging performance compares to that of human experts. It seeks to determine whether the framework, when integrated with LLM models, can achieve comparable or superior results to human experts in identifying and fixing errors in code.

\begin{table*}
\vspace{-15pt}
    \sffamily
    \small
    \centering
     \caption{The syntax and function error debugging with different LLMs. 
     The highest FR of each module is marked.}
     \label{Table:Model}
    \resizebox{0.99\linewidth}{!}{%
    \begin{tabular}{c| c c | c c | c c | c c| c c|c c }
    \toprule
    \multirow{2}{*}{\centering\textbf{Types}} & \multicolumn{2}{c|}{\centering\textbf{GPT-3.5}} & \multicolumn{2}{c|}{\centering\textbf{GPT-4}} & \multicolumn{2}{c|}{\centering\textbf{GPT-3.5+Knowledge}}& \multicolumn{2}{c|}{\centering\textbf{GPT-4+Knowledge}}& \multicolumn{2}{c|}{\centering\textbf{GPT-3.5+\approach}}& \multicolumn{2}{c}{\centering\textbf{GPT-4+\approach}}\\
    &  Syntax& Func.& Syntax& Func.& Syntax& Func.& Syntax &  Func.& Syntax &  Func.& Syntax &  Func.\\
    \midrule

    accu& 57.14\% & 36.67\%& 28.57\% & 60.00{\cellcolor[HTML]{88dc55}}\%& 47.61\% & 33.33\%& 66.67\%& 40.00\% &42.86\% &33.33\% &74.29{\cellcolor[HTML]{55dcc3}}\% &50.00\%\\ \midrule
    adder\_8bit& 62.50\% & 58.33\%& 91.67\% & 91.67\%& 50.00\% & 58.33\%& 100.00\%& 100.00\% &66.67\% &91.67\% &100.00{\cellcolor[HTML]{55dcc3}}\% &100.00{\cellcolor[HTML]{88dc55}}\%\\ \midrule
    adder\_32bit& 62.96\% & 46.67\%& 85.19\% & 66.67\%& 51.85\% & 33.33\%& 88.89\%& 73.33\% &77.78\% &46.67\% &97.14{\cellcolor[HTML]{55dcc3}}\% &94.00{\cellcolor[HTML]{88dc55}}\%\\ \midrule
    adder\_pipe\_64bit& 23.81\% & 40.00\%& 33.33\% & 26.67\%& 23.81\% & 40.00\%& 95.24{\cellcolor[HTML]{55dcc3}}\%& 86.67\% &42.86\% &60.00\% &94.29\% &90.00{\cellcolor[HTML]{88dc55}}\%\\ \midrule
    div\_16bit& 20.00\% & 0.00\%& 27.78\% & 40.00\%& 16.67\% & 20.00\%& 72.22\%& 26.67\% &16.67\% &20.00\% &81.67{\cellcolor[HTML]{55dcc3}}\% &62.00{\cellcolor[HTML]{88dc55}}\%\\ \midrule
    multi\_booth\_8bit& 100.00\% & 26.67\%& 75.00\% & 33.33\%& 100.00\% & 46.67\%& 100.00\%& 80.00{\cellcolor[HTML]{88dc55}}\% &100.00\% &60.00\% &100.00{\cellcolor[HTML]{55dcc3}}\% &77.50\%\\ \midrule
    multi\_pipe\_8bit& 9.52\% & 33.33\%& 80.95\% & 73.33\%& 28.57\% & 46.67\%& 100.00\%& 60.00\% &28.57\% &42.86\% &100.00{\cellcolor[HTML]{55dcc3}}\% &80.00{\cellcolor[HTML]{88dc55}}\%\\ \midrule
    radix2\_div& 74.07\% & 61.11\%& 11.11\% & 50.00\%& 66.67\% & 55.56\%& 70.83\%& 55.56\% &75.00{\cellcolor[HTML]{55dcc3}}\% &61.11{\cellcolor[HTML]{88dc55}}\% &66.25\% &46.00\%\\ \midrule
    alu& 28.57\% & 60.00\%& 61.90\% & 86.67\%& 66.67\% & 66.67\%& 90.48\%& 93.33\% &85.71\% &73.33\% &97.14{\cellcolor[HTML]{55dcc3}}\% &97.50{\cellcolor[HTML]{88dc55}}\%\\ \midrule
    asyn\_fifo& 37.50\% & 37.50\%& 83.33\% & 54.16\%& 25.00\% & 33.33\%& 91.67{\cellcolor[HTML]{55dcc3}}\%& 62.50\% &41.67\% &33.33\% &89.29\% &78.57{\cellcolor[HTML]{88dc55}}\%\\ \midrule
    freq\_div& 100.00\% & 83.33\%& 100.00\% & 100.00\%& 95.24\% & 83.33\%& 100.00\%& 83.33\% &100.00\% &94.44\% &100.00{\cellcolor[HTML]{55dcc3}}\% &100.00{\cellcolor[HTML]{88dc55}}\%\\ \midrule
    parallel2serial& 45.83\% & 50.00\%& 75.00\% & 75.00\%& 66.67\% & 91.67\%& 91.67\%& 100.00\% &66.67\% &100.00\% &95.00{\cellcolor[HTML]{55dcc3}}\% &100.00{\cellcolor[HTML]{88dc55}}\%\\ \midrule
    serial2parallel& 75.00\% & 26.67\%& 79.17\% & 40.00\%& 75.00\% & 20.00\%& 100.00{\cellcolor[HTML]{55dcc3}}\%& 26.67\% &79.17\% &20.00\% &98.75\% &58.75{\cellcolor[HTML]{88dc55}}\%\\ \midrule
    traffic\_light& 19.04\% & 20.00\%& 0.00\% & 40.00\%& 9.52\% & 46.67{\cellcolor[HTML]{88dc55}}\%& 90.48\%& 40.00\% &28.57\% &40.00\% &95.71{\cellcolor[HTML]{55dcc3}}\% &44.00\%\\ \midrule
    width\_8to16& 73.33\% & 74.44\%& 100.00\% & 100.00\%& 100.00\% & 80.00\%& 100.00\%& 80.00\% &100.00\% &100.00{\cellcolor[HTML]{88dc55}}\% &100.00{\cellcolor[HTML]{55dcc3}}\% &97.50\%\\
    \bottomrule
     \emph{FR}& 54.28\% & 45.73\%& 62.83\% & 61.96\%& 55.26\% & 49.57\%& 90.69\%& 66.24\% &64.26\% &56.90\%&\textbf{92.68{\cellcolor[HTML]{55dcc3}}\%} &\textbf{78.39{\cellcolor[HTML]{88dc55}}\%}\\
    \bottomrule
    \end{tabular}
    }
\end{table*}



\vspace{-2.5pt}
\subsection{Evaluation Metrics}
\vspace{-2.5pt}
\label{sbsc:EM}

\parlabel{Fix Rate (FR).} In recent work, such as ~\cite{chen2021evaluating,fakhoury2024llmbased}, the use of pass@k metrics to assess function correctness was mentioned. 
For each problem in the problem set, k code samples are generated at a time, and the problem is considered solved if any of the k samples pass the simulation test (without syntax and function errors).

Specifically, we used Fix Rate (FR) to quantify the debug ability of the debugging framework\cite{tian2024debugbench}. 
For an error code $\theta_{i}$ and its fixed version $\theta_{i}^{*}$, we had a corresponding set of test cases in testbench $\left(x_{i}^{0}, y_{i}^{0}\right)$,$\left(x_{i}^{1}, y_{i}^{1}\right)$, \ldots,$\left(x_{i}^{m}, y_{i}^{m}\right) $. 
For the correct version of the code, $\theta_{i}^{*}$, it should produce the correct output $y_{i}^{j}$ when applied to the input data $x_{i}^{j}$ from the test cases. 
That is, $a_{\theta_{i}^{*}}\left(x_{i}^{j}\right)=y_{i}^{j}$, the test case $\left(x_{i}^{j}, y_{i}^{j}\right) $ can be regarded as passing. 
Whether the error is successfully fixed can be described as $\bigwedge_{j=0}^{m}\left[a_{\theta_{i}^{*}}\left(x_{i}^{j}\right)=y_{i}^{j}\right]$, an aggregate result of all test cases.  
The FR that represents the test result on the bug instances are defined as:

\begin{equation}
\label{eq:fixrate}
\text { \textbf{FR} }=\sum_{i=0}^{n} \frac{\bigwedge_{j=0}^{m}\left[a_{\theta_{i}^{*}}\left(x_{i}^{j}\right)=y_{i}^{j}\right]}{n} \times 100 \%
\end{equation}

It is worth noting that all FR presented in this paper are calculated based on the average of 10 repeated experiments. 





\parlabel{Execution time.} This paper also considers the execution time of the framework as an important indicator of the performance, which is determined as the time elapsed between when the \name\ receives the initial design files and \name\ outputs the final modified code.


\begin{figure}[t]
    \centering
    \subfigure[FR of syntax and function errors.]
    {
    \centering
    \label{fig:tem1}    \includegraphics[width=0.475\columnwidth]{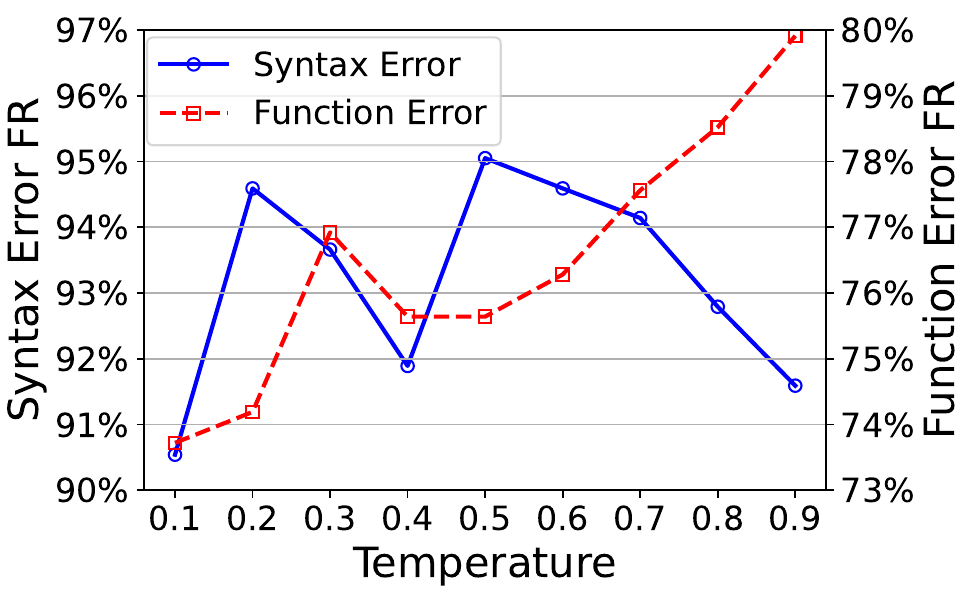}
    }
    \subfigure[Average FR.]{
    \centering
    \includegraphics[width=0.42\columnwidth]{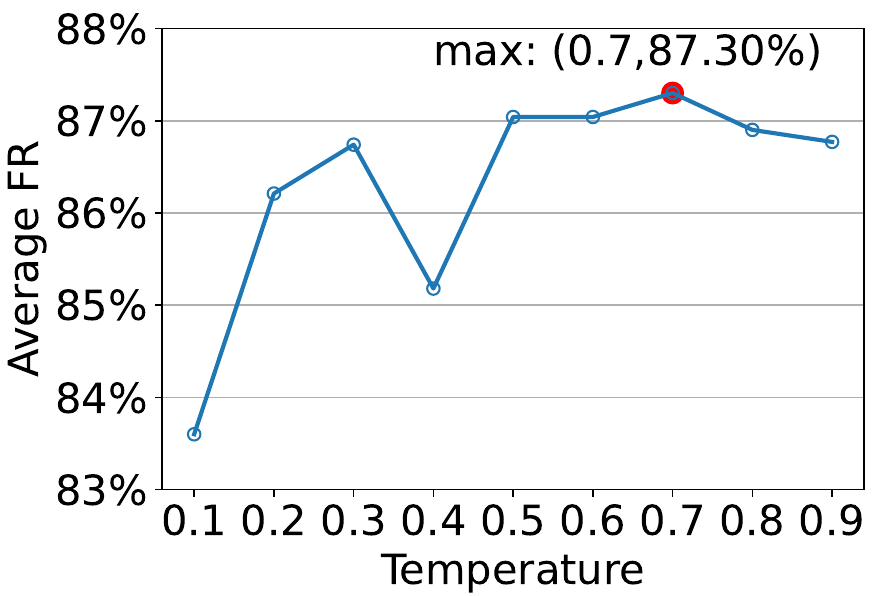}
    \label{fig:tem2}
    }
    \vspace{-5pt}
    \caption{FR against the temperature of the debug agent.}
    \vspace{-3pt}
    \label{fig:tem}
\end{figure}

\subsection{Results and Discussions}
\label{sbsc:eva}

\noindent \textbf{RQ1 (Sensitivity).}
Figure~\ref{fig:tem1} illustrates the impact of temperature on the FR of syntax errors and function errors. 
The results indicate that within the temperature range of the experiment, higher temperatures result in higher function error FR and lower syntax error FR. The function and syntax error FR reach their highest point of 80\% and 95\% respectively with temperature settings being 0.9 and 0.5 respectively.
This may be attributed to the fact that syntax error is relatively straightforward, whereas function error is more complex.  
While attempting to rectify a syntax error, the agent with a higher temperature may inadvertently introduce alterations that result in the generation of new errors.
Such errors were rarely corrected by the debug agent in the subsequent iterations according to our observation. 
In the case of function errors, the higher degree of randomness allowed LLM to to avoid modifying the same error all the time.  
Based on Figure~\ref{fig:tem1}, we calculated an average FR of all our test cases as shown in Figure~\ref{fig:tem2}. 
The overall FR reached the best case (87.30\%) when the temperature was 0.7. 

\begin{figure}[t]
    \centering    \hspace{-10pt}\includegraphics[width=1.0413\linewidth]{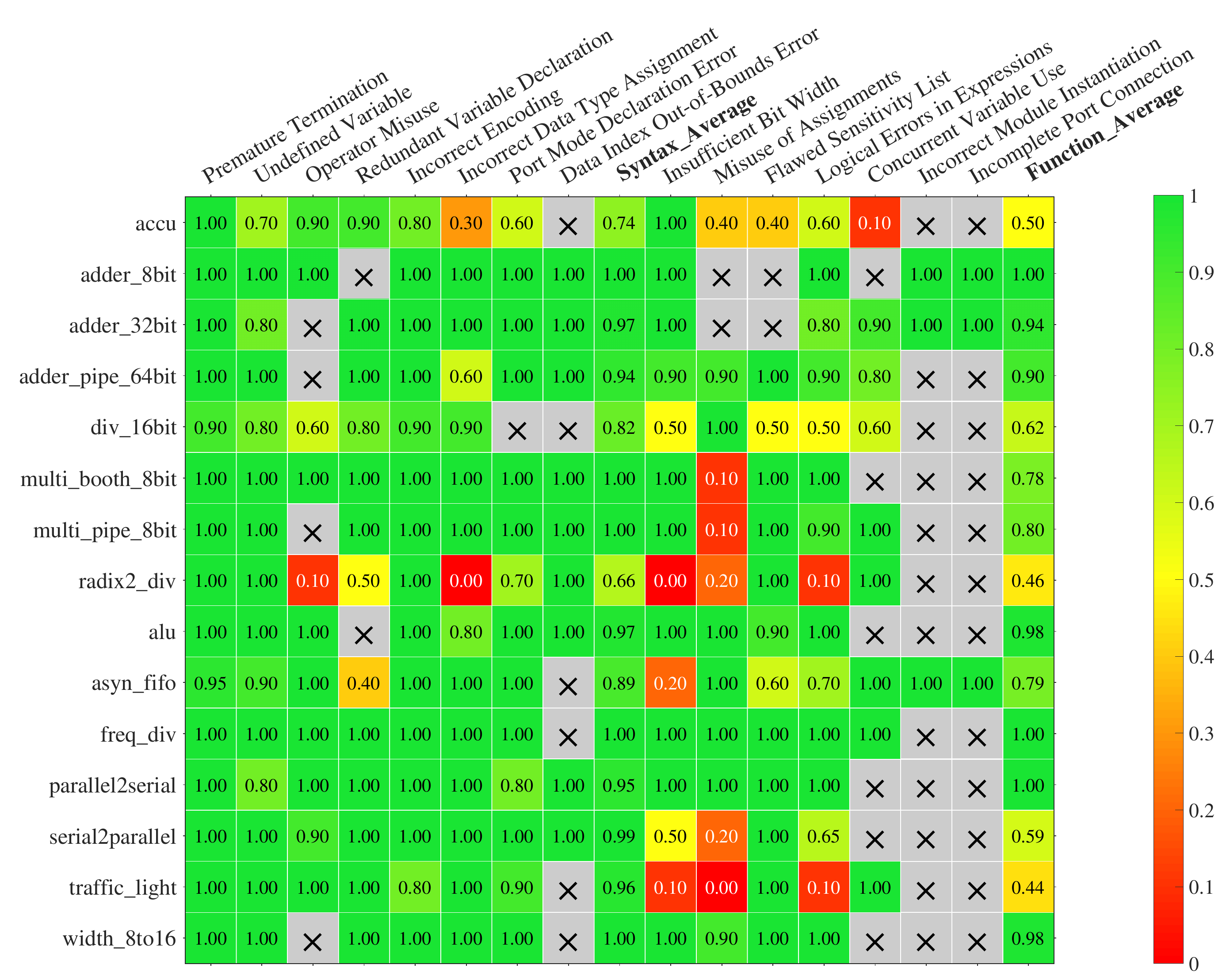}
    \caption{Heatmap result for FR. The symbol X represents an error that could not be imposed due to the limitations of the specific module structure. Syntax\_Average and Function\_Average represent the arithmetic mean of the FR for syntax errors and function errors, respectively.}
    \label{fig:rlt1}
\end{figure}

\noindent \textbf{RQ2 (Effectiveness).} Figure~\ref{fig:rlt1} shows the \approach's FR for 8 syntax errors and 7 function errors across 15 common hardware modules. 
The FR was calculated based on Equation~\ref{eq:fixrate}, and the values were colouring-coded for readability. 
The results suggest that the FR varied significantly depending on module complexity and error types. 
For instance, for modules with straightforward logic and shorter lengths such as the \emph{adder\_8bit} module, \approach\ consistently achieved a high FR, indicating its effectiveness in correcting all error types. 
Conversely, for more intricate modules like \emph{accu}, the FR diminished, highlighting the challenge of debugging such code. 
Regarding error types, while syntax errors exhibited a higher (10\% higher) overall FR than function errors, the latter posed greater difficulty in correction, particularly in complex modules. 
On average, the \approach\ achieved FR of 93\% for syntax errors and 78\% for function errors, demonstrating a greater effectiveness than existing practices. 
By contrast, the average FR achieved by RTLFixer~\cite{tsai2023rtlfixer} stood at 16\%.

\noindent \textbf{RQ3 (Impactability).} Table~\ref{Table:Model} compares the FR of both syntax and function errors achieved by two LLM models (GPT-3.5 and GPT-4) in their standard forms, after incorporating external domain-specific knowledge and with the integration of \approach. 
Results indicate that, for the same core GPT models, the models integrating \approach\ achieved the highest FR of both syntax and function errors, followed by the models incorporating knowledge only. 
The standard form models exhibited the lowest FR.

\begin{table}
    \sffamily
    \small
    \centering
     \caption{ Execution time of \approach\ against human (s: syntax error; f: function error; "Total" is calculated in seconds).}
     \label{Table:runtime2}

     \begin{threeparttable}
     \resizebox{1\linewidth}{!}{%
    \begin{tabular}{c|  p{0.5cm}  p{0.5cm} p{0.5cm}  p{0.5cm} p{0.5cm}|  c|c }
    \toprule
    \multirow{2}{*}{\centering\textbf{Types}} & \multicolumn{5}{c|}{\centering\textbf{GPT-4+\approach}} &{\centering \textbf{Human}}&    \multirow{2}{*}{\centering\textbf{Speedup}} \\
    &   Simu.& Debug& Score& Trans.& Total& Total& \\
    \midrule
    accu s &  4.4\%& 88.0\%& 5.6\% &2.1\% &116.0 &382 &3.29x \\ \midrule
    adder\_8bit s&  8.0\%& 82.6\%& 6.0\% &3.4\% &30.6 &136 &4.44x \\ \midrule
    adder\_32bit s&  2.4\%& 93.4\%& 2.4\% &1.9\% &130.3 &402 &3.09x \\ \midrule
    adder\_pipe\_64bit s&  1.8\%& 94.4\%& 2.5\% &1.4\% &169.4 &575 &3.39x \\ \midrule
    div\_16bit s&  6.9\%& 82.9\%& 8.5\% &1.7\% &63.6 &249 &3.91x \\ \midrule
    multi\_booth\_8bit s&  9.5\%& 77.6\%& 9.0\% &3.9\% &23.6 &272 &\textbf{11.53x} \\ \midrule
    multi\_pipe\_8bit s&  4.2\%& 90.1\%& 3.9\% &1.8\% &53.2 &775 &\textbf{14.56x} \\ \midrule
    radix2\_div s&  3.0\%& 89.1\%& 4.1\% &3.8\% &219.9 &620 &2.82x \\ \midrule
    alu s& 3.4\%& 90.5\%& 3.5\% &2.6\% &80.6 &318 &3.95x \\ \midrule
    asyn\_fifo s&  2.2\%& 93.4\%& 3.0\% &1.4\% &146.6 &827 &5.64x \\ \midrule
    freq\_div s&  8.3\%& 81.3\%& 6.9\% &3.5\% &25.9 &197 &7.60x \\ \midrule
    parallel2serial s&  8.2\%& 80.6\%& 8.6\% &2.6\% &32.9 &239 &7.26x \\ \midrule
    serial2parallel s&  8.8\%& 79.2\%& 8.1\% &3.8\% &25.5 &268 &\textbf{10.49x} \\ \midrule
    traffic\_light s&  3.9\%& 77.9\%& 3.9\% &14.4\% &68.3 &284 &4.16x \\ \midrule
    width\_8to16 s&  8.8\%& 80.3\%& 7.5\% &3.3\% &24.3 &232 &9.56x \\ \midrule
    accu f&  4.6\%& 87.5\%& 5.9\% &2.0\% &223.4 &1578 &7.06x \\ \midrule
    adder\_8bit f&  9.1\%& 81.9\%& 6.3\% &2.7\% &28.8 &293 &\textbf{10.17x} \\ \midrule
    adder\_32bit f&  2.1\%& 92.8\%& 2.4\% &2.8\% &168.7 &871 &5.16x \\ \midrule
    adder\_pipe\_64bit f&  1.9\%& 94.1\%& 2.3\% &1.7\% &205.1 &1814 &8.84x \\ \midrule
    div\_16bit f&  5.3\%& 83.1\%& 7.6\% &4.0\% &145.3 &1482 &\textbf{10.20x} \\ \midrule
    multi\_booth\_8bit f&  5.9\%& 83.2\%& 8.4\% &2.5\% &97.9 &816 &8.33x \\ \midrule
    multi\_pipe\_8bit f&  3.0\%& 92.1\%& 3.6\% &1.4\% &204.1 &915 &4.48x \\ \midrule
    radix2\_div f&  2.9\%& 92.0\%& 4.3\% &0.8\% &427.5 &1650 &3.86x \\ \midrule
    alu f&  3.7\%& 91.5\%& 3.6\% &1.2\% &85.8 &939 &\textbf{10.94x} \\ \midrule
    asyn\_fifo f&  1.7\%& 92.5\%& 3.1\% &2.7\% &331.7 &1746 &5.26x \\ \midrule
    freq\_div f&  8.6\%& 81.9\%& 6.8\% &2.7\% &31.8 &1527 &\textbf{48.00x} \\ \midrule
    parallel2serial f&  11.9\%& 75.0\%& 9.4\% &3.7\% &22.0 &677 &\textbf{30.80x} \\ \midrule
    serial2parallel f&  5.6\%& 77.7\%& 7.6\% &9.1\% &183.7 &993 &5.41x \\ \midrule
    traffic\_light f&  2.8\%& 84.6\%& 4.0\% &8.6\% &497.3 &1869 &3.76x \\ \midrule
    width\_8to16 f&  9.1\%& 79.6\%& 8.4\% &2.9\% &34.1 &912 &\textbf{26.74x} \\ 
    \bottomrule
    Average &  3.6\%& 88.4\%& 4.5\% &3.5\% &129.9 &795 &\textbf{6.12x} \\
    \bottomrule
    \end{tabular}
    }
     \end{threeparttable}    
\end{table}

Furthermore, when comparing the best performance (colouring-coded) across all LLMs and their variants, the models with \approach\ accounted for 80\%  of the best results, whereas the models with knowledge and standard-form models only scored 16\% and 4\%, respectively. 
It is worth noting that standard models benefited significantly from adding knowledge for identifying and correcting syntax errors, but this was less effective for function errors. 
With the integration of \approach, while improvements were observed in both syntax and function error debugging, the performance gap is more favourable for function errors, as GPT-4 with \approach\ improved the FR to 78.30\% from 66.24\% for GPT-4 with knowledge, representing an improvement of over 12\%. 
This result suggested that the integration of \approach\ could indeed enhance the performance of standard models, surpassing those with incorporating knowledge.

\noindent \textbf{RQ4 (Usability).} According to Table~\ref{Table:Model}, GPT-4 consistently outperformed GPT-3.5 across their standard forms. 
After incorporating external knowledge, the performance of GPT-3.5 still failed to meet the standard form of GPT-4. 
This demonstrated differences in the debugging capabilities of the models themselves. 
However, after integrating \approach, GPT-3.5 achieved FR of 64\% for syntax errors and 56\% for function errors, the performance was comparable to or even exceeds the standard form of GPT-4, highlighting the framework's effectiveness in directing LLM models' debugging capability.

\noindent \textbf{RQ5 (Performability).} To assess the proposed framework's effectiveness compared to human experts, we compared their debugging performance across various modules and error types as shown in Table~\ref{Table:runtime2}. 
While human experts are experienced in debugging, the framework demonstrated competitive performance in addressing syntax errors and modules with simple logic. 
For example, in the \textit{ multi\_pipe\_8bit} module, \approach\ had a 14.56x speedup. 
This performance gap was further increased for more complex function errors as \approach\ demonstrated up to 48x speedup of the human expert. 
This result illustrated the significant enhancement in the debugging capabilities with greater automation and improved efficiency.

\vspace{-2.5pt}
\section{Related Work}
\label{sc:RelatedWork}
\vspace{-2.5pt}

Recent advances in LLMs have significantly transformed hardware design, primarily through enhanced efficiency and automation~\cite{liu2023verilogeval,thakur2023verigen,blocklove2023chip, delorenzo2024make}. A key application of these models is in RTL debugging, which represents a substantial portion of total design costs. In response, various approaches, such as RTLFixer~\cite{tsai2023rtlfixer}, SBF~\cite{ahmad2023fixing}, LLM4SecHW~\cite{fu2023llm4sechw}, HDLdebugger~\cite{yao2024hdldebugger}, and AssertLLM~\cite{fang2024assertllm}, have been developed to reduce costs and increase efficiency in this area. 

These existing approaches have focused mainly on refining LLM models' performance by employing techniques like prompt engineering~\cite{white2023prompt,sahoo2024systematic}, model tuning~\cite{liu2023verilogeval,chang2023chipgpt}, and model training~\cite{liu2023chipnemo,goh2024english}. Although these efforts have led to some improvements, they have not yet successfully addressed applications to correcting function errors~\cite{ tian2024debugbench} nor achieved sufficient performance as measured by the \emph{pass@k} rates~\cite{ tsai2023rtlfixer}. In contrast, our approach adopts a collaborative process, by utilising two LLM models iteratively, to enhance debugging effectiveness for syntax and function errors.

\vspace{-2.5pt}
\section{Conclusion}
\vspace{-2.5pt}
\label{sc:Conclusion}
In this work, a systematical automated debugging framework, \textbf{\approach}, is introduced. The framework demonstrates that it is feasible to employ the LLMs for the purpose of debugging Verilog code, encompassing both syntax and function errors. 
The utilisation of prompt engineering and feedback engineering leads to an improvement in the debug capability of the LLMs, achieving fix rate of 93\% for syntax errors and 78\% for function errors. In comparison to human engineers, debugging with our framework has the potential to save up to 48 times the time overhead. Our work not only rethinks the Verilog code debugging process with the LLMs, but also paves the way for more efficient hardware design. 

\parlabel{Lessons we learnt.}
Throughout this study, we observed considerable variations in performance of the different LLMs when it comes to debugging RTL code. In line with findings from existing literature, it is clear that no single model can effectively manage all debugging scenarios. In addition, despite prompt engineering, model tuning, and model training bringing overall improvement to the model performance, decreased performance was observed in certain tests compared to the models in their standard forms. This observation highlights the need for setting up realistic expectations before LLM deployment and for understanding their operational limits, both of which remain an open challenge.

\clearpage
\bibliographystyle{ACM-Reference-Format}
\bibliography{ref}

\end{document}